\documentclass[UTF8,a4paper]{article}
\usepackage[UTF8, fontset=windows]{ctex}
\usepackage[sort&compress,numbers]{natbib}
\usepackage{booktabs}
\usepackage{geometry}
\usepackage{titlesec}
\usepackage{url}
\usepackage[font={small,bf}]{caption}
\usepackage[yyyymmdd]{datetime}
\usepackage{graphicx}
\usepackage{multirow}
\usepackage{amsmath}
\usepackage{lmodern}
\usepackage{hyperref}

\titleformat*{\section}{\songti\zihao{4}\bfseries}
\titleformat*{\subsection}{\heiti\zihao{5}\bfseries}
\titleformat*{\subsubsection}{\kaishu\zihao{5}\bfseries}

\newenvironment{csmtAbstract}{\noindent \kaishu \small {\bfseries Abstract:}}{}
\newenvironment{keywords}{\small \noindent{\bfseries Key Words:}}{}

\newcommand*{\affmark}[1][*]{\textsuperscript{#1}}

\makeatletter
\renewcommand\@maketitle{%
	\hfill
	\begin{minipage}{0.95\textwidth}
		\vskip 2em
		\let\footnote\thanks
		{\centering \bfseries \zihao{3} \@title \par} 
		\vskip 1em
		{\centering \zihao{5} \textbf{\@author} \par}
	\end{minipage}
	\vskip 1em \par
}
\makeatother

\title{Semi-Supervised Self-Learning Enhanced Music Emotion Recognition}

\author{
    \footnotesize
	\textmd{\large Yifu Sun\affmark[1],  Xulong Zhang\affmark[2], Monan Zhou\affmark[3], Wei Li\affmark[1,4\footnote{Corresponding Author: Wei Li (1970-), Male, Professor, weili-fudan@fudan.edu.cn
    \\  Acknowledgement: This work was jointly supported by NSFC (62171138), Key Laboratory of Intelligent Processing Technology for Digital Music (Zhejiang Conservatory of Music), Ministry of Culture and Tourism (Grant Number 2022DMKLB002). 
    }]}\\\vspace{0.27cm}
	\textmd{\affmark[1] School of Computer Science and Technology, Fudan University, Shanghai, China}\\
	\textmd{\affmark[2] Ping An Technology Co., Ltd., Shenzhen, China}\\
	\textmd{\affmark[3] Department of Music AI and Information Technology, Central Conservatory of Music, Beijing, China}\\\vspace{-0.17cm}
    \textmd{\affmark[4] Shanghai Key Laboratory of Intelligent Information Processing, Fudan University, Shanghai, China}
}

\begin{document}

\maketitle
\vspace{0.67cm}
\begin{csmtAbstract}
	Music emotion recognition (MER) aims to identify the emotions conveyed in a given musical piece. However, currently, in the field of MER, the available public datasets have limited sample sizes. Recently, segment-based methods for emotion-related tasks have been proposed, which train backbone networks on shorter segments instead of entire audio clips, thereby naturally augmenting training samples without requiring additional resources. Then, the predicted segment-level results are aggregated to obtain the entire song prediction. The most commonly used method is that the segment inherits the label of the clip containing it, but music emotion is not constant during the whole clip. Doing so will introduce label noise and make the training easy to overfit. To handle the noisy label issue, we propose a semi-supervised self-learning (SSSL) method, which can differentiate between samples with correct and incorrect labels in a self-learning manner, thus effectively utilizing the augmented segment-level data. Experiments on three public emotional datasets demonstrate that the proposed method can achieve better or comparable performance.
\end{csmtAbstract}


\section{Introduction}
\noindent
The music emotion recognition (MER) task aims to recognize the emotion expressed in a given music clip automatically. MER can be widely used in many fields, such as dynamically generating music to adapt to the emotion of scenes in movies or games \cite{gomez2021music}, music-assisted psychological or physical therapy, personalized recommendation in stream media, human-machine interaction, music retrieval, and so on, which has broad application prospects. In recent years, as the amount of data grows, data-driven deep learning methods have become the mainstream method in the music information retrieval (MIR) field \cite{zhang2021singer,sun2022investigation}.

At present, the duration of audio clips in public music emotion datasets is $30 \sim 45$ seconds. Although the longer the duration is, the more helpful it is to distinguish emotions, according to the study of music psychology, it is found that the duration of about one second of music is sufficient to evoke an emotional reaction \cite{bigand2005multidimensional}.
To address the issue of limited annotated data in emotion recognition tasks, some segment-based methods \cite{han2014speech,mao2019deep,sarkar2020recognition, he2022music} have been proposed recently, which naturally increase the amount of training data and can make full use of every audio sample in the dataset.

After the audio clip is divided into segments, Sarkar \textit{et al.} \cite{sarkar2020recognition} make each segment inherit the label of the clip containing it, which is also the simplest method, then majority vote and maximum run length are used to obtain clip-level results. However, the emotion of the music is not constant. Therefore, each segment may actually carry different emotions, which also introduces the problem of a noisy label.
He \textit{et al.} \cite{he2022music} used an unsupervised method, i.e. using autoencoder to reconstruct the masked mel spectrogram of the segment to obtain audio segment embedding. Then a supervised learning structure using Bi-directional Long Short-Term Memory (BiLSTM) is employed to capture temporal music information and perform emotion classification. However, it is unknown how many emotion-related features are included in the embedding.
In the field of speech emotion recognition, Mao \textit{et al.} \cite{mao2021enhancing} leveraged a self-learning framework to update model parameters and segment labels iteratively in the training process and used soft labels instead of hard labels, which to some extent solved the problem of noisy labels. However, only using the output probability distribution of the model as the soft label of the next epoch will excessively rely on the prediction ability of the model. Once the model makes a prediction error, this error will deepen with the training, which is called confirmation bias.

Inspired by \cite{tanaka2018joint} when the mixture of correct and incorrect labels are fed to the deep neural networks (DNNs), networks have a tendency to learn the latter after the former. Therefore, we propose a semi-supervised self-learning (SSSL) framework,
to model each training sample's loss value, and to distinguish the training samples most likely to be clean from those most likely to be noisy.
Then we use the mixup \cite{DBLP:conf/iclr/ZhangCDL18} data augmentation algorithm and consistency regularization to prevent the confirmation bias of the model's prediction. After obtaining a label noise-robust segment-level emotion predictor, we can use it to generate the predicted probabilities of each segment in the song-level data as a structured feature representation. Finally, a second machine learning algorithm is employed to predict the overall emotion for each song. Our main contributions are:
\begin{enumerate}
	\item Instead of directly inheriting clip-level labels for each segment or unsupervised methods that do not use labels at all, we use semi-supervised learning to deal with noisy labels.
	\item Combining noisy label processing with semi-supervised learning, to mitigate the confirmation bias issue associated with self-training, which can lead to the accumulation of model errors.
	\item Compared with baseline models, the effect is improved.
\end{enumerate}

\section{Method}
\begin{figure*}[bhtp]
	\centering
	\includegraphics[scale=1.0,width=12cm]{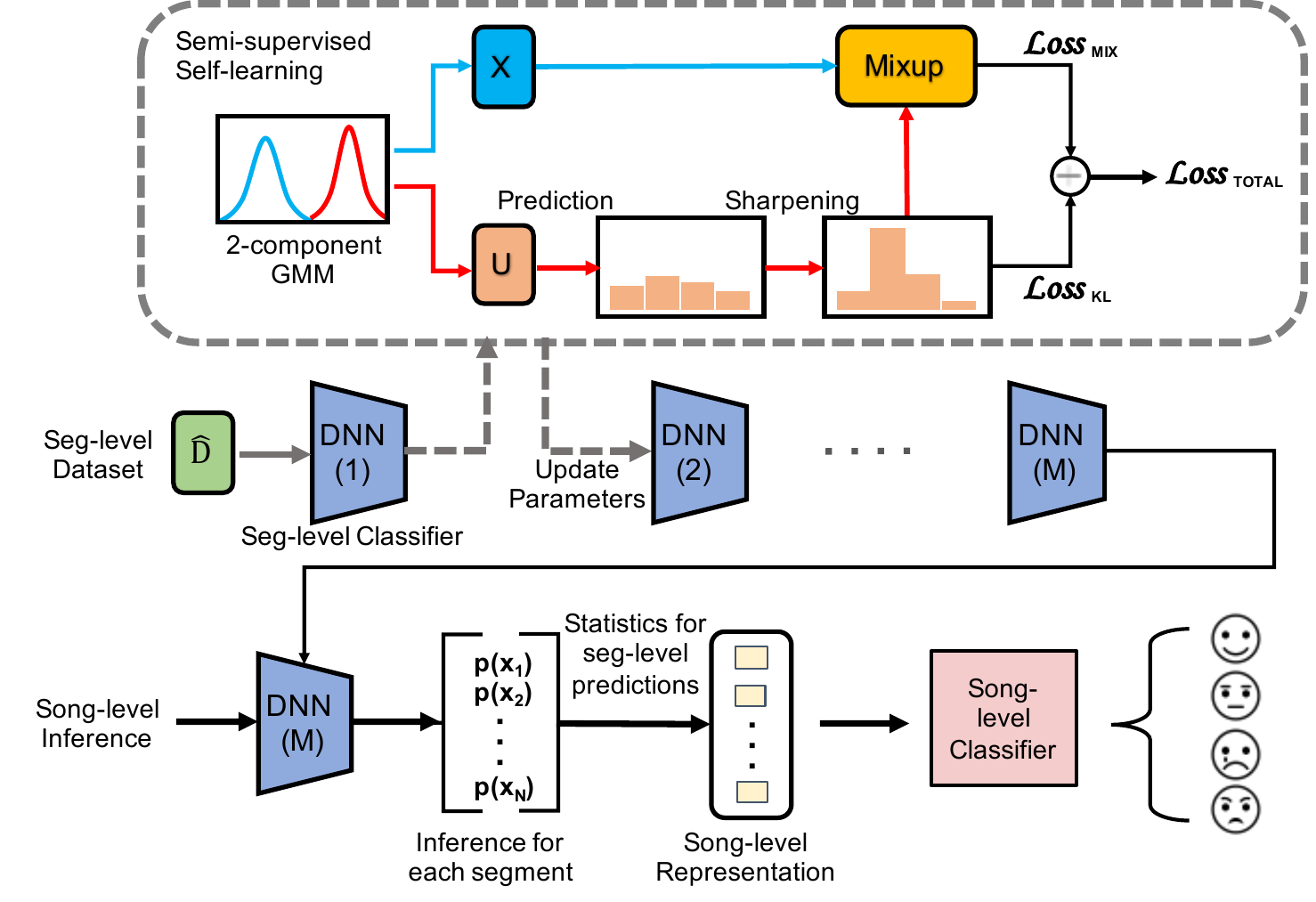}
	\caption{Flow chart of the proposed method: blue arrows indicate the clean set (Gaussian distribution with smaller mean), red arrows indicate the noisy set, which is treated as the unlabeled set in subsequent semi-supervised learning where the clean set serves as the labeled set}
	\label{fig:flow}
\end{figure*}

\noindent
Our method is divided into two steps. The first step is to train a segment-level classifier robust to label noise on the expanded segment-level dataset.
The second step uses the original song-level dataset to predict each segment in each song, obtain the statistical value of the probability distribution of each segment, and aggregate them as the feature representation of the song. Then, a machine learning method is used to complete the emotion prediction of the song.
Figure \ref{fig:flow} illustrates the overall framework of our proposed algorithm.

\subsection{SSSL Framework}
\noindent
We propose an SSSL on the extended segment dataset, aiming to obtain a label-noise robust segment-level classifier. Unlike conventional semi-supervised learning methods, our labeled data and unlabeled data generation are dynamic. Moreover, we do not assign a fixed pseudo-label to the unlabeled data.

At the beginning of each epoch, the training set is partitioned into a clean set and a labeled noisy set using a two-component Gaussian mixture model (GMM) by the cross-entropy (CE) loss value for each training sample. Then the expectation-maximization (EM) algorithms are leveraged to fit the GMM to the observation. Then, the semi-supervised learning method is used to treat the clean set as the labeled set $\mathcal{X}$ while the noisy set is the unlabeled set $\mathcal{U}$. We erase the label of the $\mathcal{U}$ set, use the predicted value of the model as the soft label, and then use a sharpening algorithm to get the pseudo soft label. Then we obtain the total loss using mixup and consistency regularization.

\subsubsection{Task Formulation}
\noindent
In a k-class classification problem, the training data with n training samples denoted as $X = [x_{1},\ldots,x_{n}]$, along with their corresponding ground-truth labels $Y = [y_{1},\ldots,y_{n}]$.
Each $y_{i}$ represents the true class label and is represented as a k-dimensional one-hot vector.
Classification problems on clean label datasets are often defined as Eq. (\ref{eq:1}), where ${\theta}$ represents the model parameters, and $\mathcal{L}$ represents the loss objective function.
\begin{equation}
	\underset{\theta}{\min} \mathcal{L}({\theta}|X,Y)
	\label{eq:1}
\end{equation}

The CE loss function in a classification problem is defined as Eq. (\ref{eq:2}), where $f$ denotes the output of the neural network classifier, i.e. the softmax layer.
\begin{equation}
	\mathcal{L} = -\frac{1}{n} \sum_{i=n}^{n}\sum_{j=1}^{k}y_{ij}log f_{j}(\theta,x_{i})
	\label{eq:2}
\end{equation}

But when we use the above formulas (\ref{eq:1}) and (\ref{eq:2}) to train on the noisy label dataset, severe overfitting will occur.

\subsubsection{Training Samples Partition}
\noindent
DNNs have been observed to prioritize learning from simple and logically consistent samples in the presence of noisy labels, resulting in reduced loss for these samples, particularly in the early stages of training \cite{zhang2021understanding}. This phenomenon suggests that the loss distributions of clean and noisy samples during training can be approximated by two Gaussian distributions, with the clean samples having a smaller mean loss. Leveraging this training characteristic, we employ a GMM to differentiate between noisy and clean samples by utilizing the per-sample loss as input.
The probability density function of the loss K-component mixture model is defined as Eq. (\ref{eq:3}), where $\lambda_{k}$ indicates the mixing coefficients for the combination of each probability density function $p(l|k)$.
\begin{equation}
	p(l) = \sum_{k=1}^{K}\lambda_{k}p(l|k)
	\label{eq:3}
\end{equation}

Regarding this case, we can utilize a two-component GMM to model the distribution of clean and noisy samples.
We utilize a two-component GMM and
then feed the CE loss $\mathcal{L}$ of every training data to the GMM. Then, to estimate the parameters of the GMM we apply the EM algorithm.
We define the posterior probability $\omega_{i}$ as the probability that the $i_{th}$ data sample belongs to the Gaussian component with a smaller mean, given its loss $l_{i}$.
By setting a global threshold $\tau$ for the probability $\omega$ of all training samples' labels to be clean, we can split the expanded segment-level dataset $\mathcal{D}$
into two parts:
set $\mathcal{X}$ and set $\mathcal{U}$.
The labels in set $\mathcal{X}$ have more possibilities to be correct, and thus, they will be used as the labeled set. On the other hand, the labels in set $\mathcal{U}$ are more likely to be incorrect, and will be erased in the subsequent steps. Therefore, set $\mathcal{U}$ will be used as the unlabeled set.

\subsubsection{Modified Semi-supervised Learning}
\noindent
We first preprocess the set $\mathcal{U}$. For the set $\mathcal{U}$,  the initial labels are likely to be incorrect and have been erased. As a result, we generate pseudo soft labels $\hat{y}$ by sharpening the predicted distribution of the model by Eq. (\ref{eq:4}), where $S(\cdot)$ indicates the temperature sharpening function often used in pseudo labeling, and $T$ represents the temperature coefficient.
\begin{equation}
	\begin{split}
		\hat{y} & = S(f_{j}(\theta,x_{i}))                                                                    \\
		        & = f_{j}(\theta,x_{i})^{-k \frac{1}{T}} / \sum_{k=1}^{K}f_{j}(\theta,x_{i})^{-k \frac{1}{T}}
	\end{split}
	\label{eq:4}
\end{equation}

And the $\mathcal{\hat{U}}$ will be generated then as Eq. (\ref{eq:5}).
\begin{equation}
	\mathcal{\hat{U}}= \{(x_{i},\hat{y_{i}}) | x_{i} \in \mathcal{U}\}
	\label{eq:5}
\end{equation}

As \cite{arazo2019unsupervised} has shown the mixup technology can eliminate confirmation bias to a certain extent.
This technique involves training on convex combinations of pairs of samples, denoted as $x_{p}$ and $x_{q}$, along with their corresponding labels $y_{p}$ and $y_{q}$ shown as Eq. (\ref{eq:7}), where $\delta$ is drawn from a beta distribution randomly.
\begin{equation}
	\begin{aligned}
		x = \delta x_{p} + (1-\delta) x_{q} \\
		y = \delta y_{p} + (1-\delta) y_{q}
	\end{aligned}
	\label{eq:7}
\end{equation}

This integration introduces regularization to encourage the network to exhibit linear behavior between samples, thereby reducing fluctuations in distant regions. In terms of label noise, mixup offers a strategy to merge clean and noisy samples, resulting in a more representative loss that guides the training procedure.

\subsubsection{Loss Function}
\noindent
Eq. (\ref{eq:7}) can be regarded in the loss as $l = \delta l_{p} + (1 - \delta) l_{q}$, and the standard CE loss is leveraged for semi-supervised learning part by Eq. (\ref{eq:8}):
\begin{equation}
	\begin{split}
		\mathcal{L}_{MIX} =
		\frac{1}{n}\sum_{i=1}^{N}\delta[y_{i,p}log(f(x_{i}))] + (1 - \delta)[y_{i,q}log(f(x_{i}))]
	\end{split}
	\label{eq:8}
\end{equation}

Confirmation bias, resulting from the accumulation of errors, is a common issue in self-training. Model ensembling is a widely adopted approach to mitigate this bias. Dropout, which can be viewed as an implicit form of model ensembling, is commonly used during network training. However, when it comes to sample selection or inference, dropout is typically disabled to ensure consistency. In the presence of label noise, the decision boundaries between classes become blurred, leading to significant inconsistencies among sub-models. To address this, we incorporate the R-Drop loss \cite{wu2021r}, a straightforward yet effective dropout regularization method shown as Eq. (\ref{eq:9}), to promote consistency among the sub-models.
\begin{equation}
	\begin{split}
		\mathcal{L}_{KL} =
		\sum_{x \in U}
		\frac{1}{2}(D_{KL}(P1(x,\theta)||P2(x,\theta)) + D_{KL}(P2(x,\theta)||P1(x,\theta)))
	\end{split}
	\label{eq:9}
\end{equation}

Therefore, the total loss is as Eq. (\ref{eq:10}), where $\lambda$ is the hyperparameter to control the weight.
\begin{equation}
	\mathcal{L} = \mathcal{L}_{MIX} + \lambda \mathcal{L}_{KL}
	\label{eq:10}
\end{equation}

\subsection{Song-level Decisions}
\noindent
The final goal is to get song-level prediction results. Given a sequence of probability distributions of emotional states generated by a segment-level classifier for a song, we can make decisions based on the information.

A reliable segment-level classifier serves as a prerequisite for clip-level classification. This allows us to utilize machine learning algorithms to handle structured features obtained from statistical properties of the segment probability distributions.
The probability of the $k_{th}$ emotion for segment $s$ is defined as $P_{s}(E_{k})$ in Eq. (\ref{eq:11}), where $C$ represents the set of all segments in a single song.
\begin{equation}
	\begin{aligned}
		f_{1}^{k} = \max_{s \in C}P_{s}(E_{k})                  \\
		f_{2}^{k} = \min_{s \in C}P_{s}(E_{k})                  \\
		f_{3 \sim 5}^{k} = Quartiles_{1 \sim 3}\{P_{s}(E_{k})\} \\
		f_{6}^{k} = \frac{1}{|C|}\sum_{s \in C}P_{s}(E_{k})     \\
		f_{7}^{k} = \frac{|P_{s}(E_{k} > \theta)|}{|C|}
	\end{aligned}
	\label{eq:11}
\end{equation}

Features $f_{1}^{k}$, $f_{2}^{k}$ and $f_{3}^{k}$ correspond to the maximum value of the segment-level probability of the $k_{th}$ emotion in the song, minimum, and average values, respectively. The features $f_{3}^{k} \sim f_{5}^{k}$ correspond to the $k_{th}$ emotion's three quartiles in the song respectively.
The feature $f_{7}^{k}$ represents the percentage of segments with a high probability of sentiment k.
The feature exhibits low sensitivity to a threshold, allowing for empirical selection.
This aggregation step produces a feature representation of dimension $7 \times K$ for each song.
With this set of song-level feature representations, we can train a secondary classifier that is relatively simple in nature to make decisions at the song level.

\section{Experiments}
\noindent
We tested our proposed algorithm on three public datasets: PMEmo dataset \cite{zhang2018pmemo}, Emotion in Music (EiM) \cite{soleymani20131000} and 4Q dataset \cite{panda2018novel}.
Since deep learning is data hungry, these data volumes are difficult to support models with strong training generalization ability and are easy to overfit.

\subsection{Experiment Setup}
\noindent
We convert each segment unit into a mel spectrogram,
with the Hanning window length of 1,024 and window hop size of 512 using the \textit{librosa} \cite{mcfee2015librosa}. We utilize 128 mel bins.
Once the audio clip is segmented into smaller parts, the amount of training samples is dozens of times larger than the original.
Our segment classifier uses VGG16 \cite{DBLP:journals/corr/SimonyanZ14a}, we only modify the input channel and output number of classification. The support vector machine (SVM) is used for song-level decisions.

\subsection{Datasets}
\label{Datasets}
\noindent \textbf{PMEmo}: This dataset contains 794 pieces of music, all of which are pop music in mp3 format. The audio duration is mainly around 45 seconds.
We utilize 767 of these clips with static valence/arousal (V/A) annotations in our work.

\noindent \textbf{4Q}: There are 900 music clips in this dataset. All music clips are divided into four parts according to Russell's V/A quadrant, with 225 clips in each part. Most audio clips are approximately 30 seconds.

\noindent \textbf{EiM}: The dataset consists of 1,000 music samples, each 45 seconds long, obtained from sources like Jamendo, with copyrighted music. The labels for this dataset were obtained through crowdsourcing platforms. Just like in previous works, we utilized a set of 744 audio samples after removing duplicates.

\subsection{Audio Pre-processing}\label{Audiopre}
\noindent
In the previous works, in the methods of taking the entire clip as input \cite{DBLP:conf/ismir/DelbouysHPRM18, de2020multiple}, 30 seconds of audio are generally reserved, and the part less than 30 seconds is padded with zeros. In \cite{he2022music}, for clips less than 30 seconds, they padded them to 30 seconds by repeating themselves continuously from the beginning to the end.

Since the variance of sample duration in different data sets is large,
when the method of filling zeros is used, many blank segments will be generated, which is invalid training data. Using the circular padding method will generate too many repeated training samples. Therefore, we did not do any padding.

\section{Results}
\noindent
In this section, we present the performance results for different segment durations and compare them with the results obtained from other MER methods.

\subsection{Performance at Different Segment Duration}
\noindent
According to the previous music psychology research \cite{bigand2005multidimensional, xiao2008best}, people can react to and make judgments about the emotions in music within 1s. The longer the segment, the more helpful it is for emotion recognition, but too long segments will reduce the data volume of the segment dataset, so a compromise is required. Thus, we experimented with integer segment duration from 1s to 5s, the overlap of adjacent segments is 1s less than the segment duration.
After segmentation, see Table \ref{tab:seg_clip} for the sample number comparison between the expanded segment-level dataset and the original song-level dataset.
The experimental results of different segment durations on binary classification are shown in Table \ref{tab:seg com}.

\begin{table}[hbt]
	\centering
	\caption{\textbf{Number of samples after segmentation}}
	\begin{tabular}{cccc}
		\toprule
		\textbf{Dataset}     & \textbf{\#Song}      & \textbf{Seg Dur.(Overlap)} & \textbf{\#Segment} \\ \hline
		\multirow{5}*{PMEmo} & \multirow{5}*{767}   & 1(0)                       & 22,945             \\
		                     &                      & 2(1)                       & 22,184             \\
		                     &                      & 3(2)                       & 21,411             \\
		                     &                      & 4(3)                       & 20,640             \\
		                     &                      & 5(4)                       & 19,869             \\
		\hline
		\multirow{5}*{4Q}    & \multirow{5}*{900}   & 1(0)                       & 26,957             \\
		                     &                      & 2(1)                       & 26,046             \\
		                     &                      & 3(2)                       & 25,135             \\
		                     &                      & 4(3)                       & 24,231             \\
		                     &                      & 5(4)                       & 23,330             \\
		\hline
		\multirow{5}*{EiM}   & \multirow{5}*{1,000} & 1(0)                       & 29,982             \\
		                     &                      & 2(1)                       & 28,981             \\
		                     &                      & 3(2)                       & 27,976             \\
		                     &                      & 4(3)                       & 26,973             \\
		                     &                      & 5(4)                       & 25,971             \\

		\bottomrule
	\end{tabular}
	\label{tab:seg_clip}
\end{table}

\begin{table}[hbt]
	\centering
	\caption{Experimental results with different segment duration}
	\begin{tabular}{cccccc}
		\toprule
		\multirow{2}*{\textbf{Datasets}} & \multirow{2}*{\textbf{Seg Dur.}} & \textbf{Valence} &                & \textbf{Arousal} &                \\ \cline{3-4} \cline{5-6}
		                                 &                                  & \textbf{Acc.}    & \textbf{F1}    & \textbf{Acc.}    & \textbf{F1}    \\
		\hline
		\multirow{5}*{PMEmo}             & 1                                & 77.43            & 81.32          & \textbf{85.42}   & \textbf{86.61} \\
		                                 & 2                                & 78.71            & 82.20          & 84.20            & 85.26          \\
		                                 & 3                                & \textbf{83.19}   & \textbf{82.31} & 84.49            & 86.01          \\
		                                 & 4                                & 82.11            & 81.71          & 81.20            & 82.20          \\
		                                 & 5                                & 82.69            & 81.32          & 84.42            & 82.61          \\
		\hline
		\multirow{5}*{4Q}
		                                 & 1                                & 69.11            & 67.32          & \textbf{87.20}   & 86.67          \\
		                                 & 2                                & 71.32            & 72.45          & 86.60            & \textbf{86.70} \\
		                                 & 3                                & \textbf{75.20}   & \textbf{77.37} & 86.56            & 86.14          \\
		                                 & 4                                & 74.32            & 76.55          & 85.30            & 86.65          \\
		                                 & 5                                & 74.59            & 77.05          & 85.67            & 86.51          \\

		\bottomrule
	\end{tabular}
	\label{tab:seg com}
\end{table}
The findings also indicate that shorter segment durations exhibit better performance in capturing the arousal dimension, whereas longer segment durations are advantageous for recognizing valence. For example, in the PMEmo dataset, the 1s segment showed the best arousal result with an accuracy of 85.42\% and an F1-score of 86.61\%, while the 3s segment showed a better valence accuracy of 83.19\% and F1-score: 82.31\%. The results on the 4Q dataset show a similar trend.

For such results, our analysis may be that arousal and intensity are more correlated, and intensity is relatively easier to be preferentially recognized by the auditory system. The identification of valence requires a longer period, this dimension involves more psychological knowledge, and the perception process of psychology is much more complicated, so it needs a longer time period \cite{bigand2005multidimensional}. Moreover, we found it is not that the longer the segment duration, the better the experimental results.
It may be related to the overlap value used by different segment durations. To ensure the amount of data, we leverage a larger overlap value in long segments, which will lead to data redundancy and make the model overfit the data. Leading to poor generalization ability to new data.

\subsection{Validation Experiment}
\begin{figure}[hbt]
	\centering
	\includegraphics[scale=1.0,width=10cm]{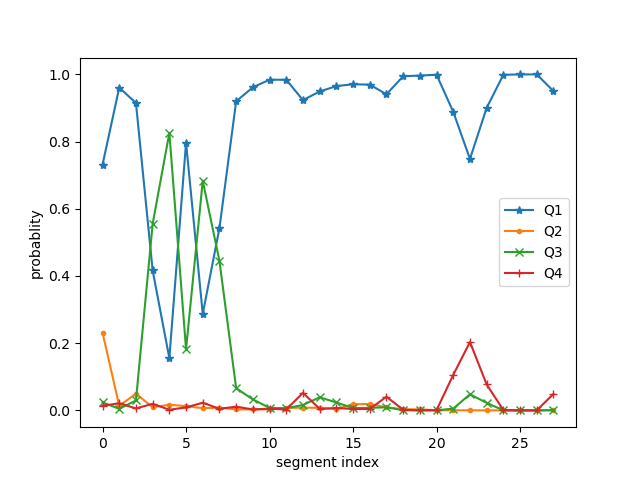}
	\caption{Segment-level classifier output on test sample MT0010465830}
	\label{fig:out}
\end{figure}

\noindent
Figure \ref{fig:out} shows an example of emotion predictions for each segment of the test sample MT0010465830 in the 4Q dataset, which is labeled as happy (Q1). The segment classifier has four outputs corresponding to four different emotional states: Q1, Q2, Q3, and Q4.

As shown in the figure, the probabilities for each emotion change throughout the entire song. The true emotion of the whole song is Q1, and as seen in the figure, the Q1 probability is highest in most segments, but there are also other emotions dominating some parts of the song. While not all songs exhibit distinctive segment-level outputs, we can employ the song-level classifier to discern them. The high-pitched and cheerful singing of the male singer at the beginning of the song kept the Q1 prediction probability the highest for the first few seconds. Then the vocals quickly changed to a low tone, and Q3 (sadness) took over. At 7s, the trumpet and drums come in and shift the emotion back to happiness. In the 20s, although the vocals turned to a low tone again, the background drumbeats and trumpet sounds continued, so the Q1 probability only slightly decreased, still dominant overall. The audio content confirms that the emotional expression in music changes dynamically throughout the song, rather than being consistent throughout.

\subsection{Experimental Results Compared with Other Models}
\noindent
In order to make a fair comparison, our experimental results are all the average values obtained under the cross-validation of ten fold.
Among them, the proposed* is the ablation experiment, which does not contain $L_{KL}$. Bold numbers indicate the best result.
\begin{table}[!hbt]
	\centering
	\caption{Comparison with other methods}
	\scalebox{0.92}{
		\begin{tabular}{cccccccc}
			\toprule
			\multirow{2}*{\textbf{Datasets}} & \multirow{2}*{\textbf{Method}}
			                                 & \multicolumn{4}{c}{\textbf{Binary classification}} & \multicolumn{2}{c}{\textbf{Four classification}}                                                                                      \\

			\cmidrule(lr){3-6} \cmidrule(lr){7-8}

			                                 &                                                    & \textbf{V-acc}                                   & \textbf{V-f1}  & \textbf{A-acc} & \textbf{A-f1}  & \textbf{Acc}   & \textbf{F1}    \\
			\hline
			\multirow{4}*{PMEmo}             & SVM \cite{yin2019user}                             & 70.43                                            & 75.32          & 71.49          & 76.36          & --             & --             \\
			                                 & 2-Stage \cite{he2022music}                         & 79.01                                            & \textbf{83.20} & 83.20          & 83.20          & --             & --             \\
			                                 & Proposed*                                          & 79.01                                            & 82.01          & 81.20          & 82.20          & --             & --             \\
			                                 & Proposed                                           & \textbf{83.19}                                   & 82.31          & \textbf{85.42} & \textbf{86.61} & --             & --             \\
			\hline
			\multirow{5}*{4Q}
			                                 & 2-Stage \cite{he2022music}                         & 67.11                                            & 67.11          & 86.56          & 86.56          & --             & --             \\
			                                 & Audio Embedding \cite{DBLP:conf/aaai/KohD21}       & --                                               & --             & --             & --             & 72.00          & --             \\
			                                 & SVM \cite{panda2018novel}                          & --                                               & --             & --             & --             & --             & 76.41          \\
			                                 & Proposed*                                          & 76.32                                            & 76.45          & 86.70          & 86.50          & 74.39          & 75.30          \\
			                                 & Proposed                                           & \textbf{77.20}                                   & \textbf{78.37} & \textbf{87.20} & \textbf{86.67} & \textbf{76.49} & \textbf{77.60} \\
			\hline
			\multirow{3}*{EiM}
			                                 & Audio Embedding \cite{DBLP:conf/aaai/KohD21}       & --                                               & --             & --             & --             & 72.00          & --             \\
			                                 & Proposed*                                          & --                                               & --             & --             & --             & 72.90          & 73.32          \\
			                                 & Proposed                                           & --                                               & --             & --             & --             & \textbf{75.81} & \textbf{75.34} \\
			\bottomrule
		\end{tabular}
	}
	\label{tab:binary class}
\end{table}

In comparison to the models listed in Table \ref{tab:binary class}, our model, which utilizes segment-level data, demonstrates superior performance when compared to other models that directly use entire music clips.
Emotional states in long music fragments may have changed or be in transition between different emotional states \cite{xiao2008best}, which may confuse the learning model and make it difficult to extract unified musical features specific to one emotion. In addition, the emotional value of music may be influenced by the processing of Western harmony, particularly the extensive use of minor keys. This processing requires more time to fully comprehend compared to the arousal dimension, which is primarily associated with the dynamic aspects of music stimulation \cite{bigand2005multidimensional}.
The best results of the same task have been achieved in most of the evaluation indicators of the binary classification problem.
Segment-based method alleviates this problem, as emotions tend to be more constant over shorter segment durations, which facilitates emotion recognition and improves learning efficiency.

\section{Conclusion}
\noindent
An SSSL framework is proposed to deal with the label noise problem caused by segment-based methods. The framework can use unlabeled data through self-learning, and use labeled data to guide the model to learn the correct feature representation, so as to effectively deal with the problem of label noise.
In the self-learning process, the model may be too confident in its own prediction results, resulting in high confidence in the prediction results, which will accumulate errors. This method addresses this issue by introducing an additional consistency regularization, which improves the generalization and robustness of the model.
Further research on song-level decision-making will be conducted in the future.

\renewcommand{\refname}{References}
\bibliographystyle{IEEEbib}
\bibliography{mybib}

\end{document}